\documentclass[12pt]{article}
\usepackage{graphicx}
\usepackage{bm}
\usepackage{subfigure}
\usepackage{amssymb}
\usepackage{ascmac}
\usepackage{url}
\usepackage{longtable}
\usepackage{booktabs}

\newcommand\pubdate{\today}

\def\napoli{Institute of Particle and Nuclear Studies\\
High Energy Accelerator Research Organization (KEK)\\
Oho 1-1, Tsukuba, Ibaraki, 305-0801, Japan}
\def\contact{\footnote{Corresponding author.\\E-mail address: yamadami@post.kek.jp (M. Yamada).}}

\def\Title#1{\begin{center} {\Large #1 } \end{center}}
\def\Author#1{\begin{center}{ \sc #1} \end{center}}
\def\Address#1{\begin{center}{ \it #1} \end{center}}

\newcommand\pubblock{\rightline{\begin{tabular}{l} 
         \pubdate  \end{tabular}}}
\newenvironment{Abstract}{\begin{quotation}  }{\end{quotation}}
\newenvironment{Presented}{\begin{quotation} \begin{center} 
             PRESENTED AT\end{center}\bigskip 
      \begin{center}\begin{large}}{\end{large}\end{center} \end{quotation}}

\newcommand{\um}{\mbox{$\mu$m}}
\newcommand{\vgs}{\mbox{$V_{\mathrm{GS}}$}}
\newcommand{\vd}{\mbox{$V_{\mathrm{D}}$}}
\newcommand{\id}{\mbox{$I_{\mathrm{D}}$}}
\newcommand{\vth}{\mbox{$V_{\mathrm{th}}$}}
\newcommand{\ileak}{\mbox{$I_{\mathrm{leak}}$}}
\newcommand{\gm}{\mbox{$g_{\mathrm{m}}$}}
\newcommand{\vbpw}{\mbox{$V_{\mathrm{BPW}}$}}
\newcommand{\ibpw}{\mbox{$I_{\mathrm{BPW}}$}}
\newcommand{\Eox}{\mbox{$E_{\mathrm{ox}}$}}
\newcommand{\sEox}{\mbox{$E^2_{\mathrm{ox}}$}}

\newcommand{\Tbox}{\mbox{$T_{\mathrm{BOX}}$}}

\newcommand{\idvg}{\mbox{$I_{\mathrm{D}}-V_{\mathrm{GS}}$}}
\newcommand{\xray}{$X$-ray}
\newcommand{\gray}{$\gamma$-ray}

\newcommand{\fnc}{$\ln(J/E^2_{\mathrm{ox}})$}
\newcommand{\invE}{$1/\Eox$}

\def\beq{\begin{equation}}
\def\eeq#1{\label{#1}\end{equation}}
\def\eeqn{\end{equation}}

\def\beqa{\begin{eqnarray}}
\def\eeqa#1{\label{#1}\end{eqnarray}}
\def\eeqan{\end{eqnarray}}

\let\bar=\overbar

\def\Dslash{\not{\hbox{\kern-4pt $D$}}}
\def\dslash{\not{\hbox{\kern-2pt $\del$}}}

\def\msb{{\bar{\ssstyle M \kern -1pt S}}}

\begin{document}
\begin{titlepage}
\pubblock

\vfill
\Title{Compensation of radiation damages for SOI pixel detector via tunneling}
\vfill
\Author{Miho Yamada\contact, Yasuo Arai and Ikuo Kurachi}
\Address{\napoli}
\vfill
\begin{Abstract}
We are developing monolithic pixel detectors based on SOI technology for high energy physics, $X$-ray applications and so on.
To employ SOI pixel detector on such radiation environments, we have to solve effects of total ionizing dose (TID) for transistors which are enclosed in a oxide layer.
The holes which are generated and trapped in the oxide layer after irradiation affect characteristics of near-by transistors due to its positive electric field.
Annealing and radiation of ultraviolet are not realistic to remove trapped holes for a fabricated detector due to  thermal resistance of components and difficulty of handling.
We studied compensation of TID effects by tunneling using a high voltage.
For decrease of trapped holes, applied high voltage to buried $p$-well which is under oxide layer to inject the electrons into the oxide layer.
In this report, recent progress of this study is shown.
\end{Abstract}
\vfill
\begin{Presented}
International Workshop on SOI Pixel Detector (SOIPIX2015), Tohoku University, Sendai, Japan\\3-6, June, 2015.
\end{Presented}
\vfill
\end{titlepage}
\def\thefootnote{\fnsymbol{footnote}}
\setcounter{footnote}{0}

\section{Introduction}

\begin{figure}[htbp]
  \centering
  \includegraphics[width=0.6\textwidth]{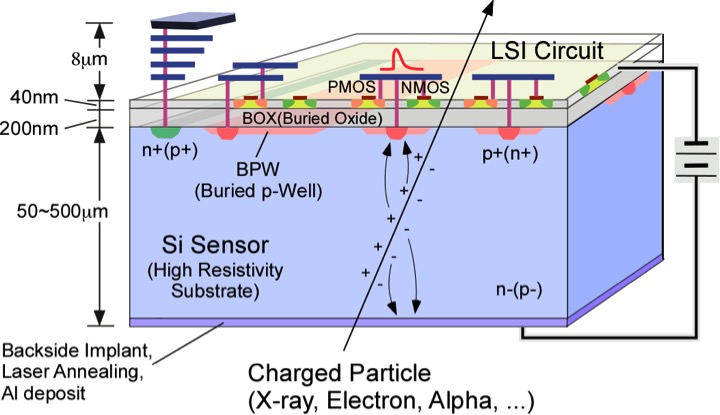}
  \caption{Cross section of SOI monolithic pixel detector.}
  \label{fig:soipix}
\end{figure}

We are developing monolithic pixel detector with Silicon-on-Insulator technology (SOIPIX) for quantum imaging in high-energy physics, astrophysics, medical imaging, material science and so on.
Cross section of the SOIPIX is shown in Figure~\ref{fig:soipix}.
LSI circuit for a signal readout is on SOI layer which is above the buried oxide layer called BOX.
High resistive substrate for sensor about $50\sim500$~\um\ is according to the application.
Bias voltage for the sensor depletion is applied between backside of the sensor and the SOI.
Buried $p$-well (BPW) which is under the BOX is to suppress the back-gate effect due to the bias voltage.

Imaging detectors used in such experiments are required to have a high tolerance for radiation.
The cause of radiation damage is trapped holes in the BOX.
Passing through charged particles, \xray\ \gray\ and so on generate the electrons and the holes in the BOX.
Since the mobility of the hole is lower than the electron, the generated holes are trapped in the BOX.
Positive potential by these trapped charges affect transistors for readout circuit.
As a compensation of such a total ionizing dose (TID) effect, annealing and radiation of UV are well known~\cite{Anneal}.
But they are very difficult particularly for inner most layer of pixel detector in high-energy accelerator experiment such as the ATLAS experiment at the LHC.

\section{Compensation of TID effect}

\begin{figure}[htbp]
  \centering
  \includegraphics[width=0.5\textwidth]{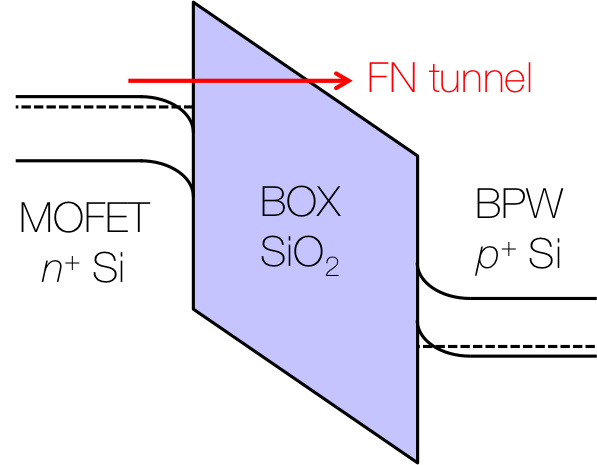}
  \caption{Fowler-Nordheim tunneling. The electrons tunnel through triangular potential of $\mathrm{Si-SiO_2}$ junction. FN tunneling occur by high electric filed around 5~MV/cm.}
  \label{fig:fn}
\end{figure}

\begin{figure}[htb]
  \centering
  \includegraphics[width=0.45\textwidth]{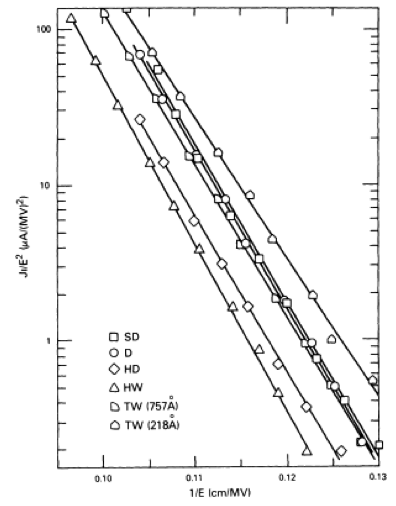}
  \caption{Fowler-Nordheim plots for variety of MOSFET samples~\cite{fnplot}. $x$-axis is \invE\ (cm/MV) and $y$-axis is \fnc\ ($\mu$A/$\mathrm{(MV)^2}$). \fnc\ as a function of \invE\ shows linear relationship.}
  \label{fig:fnplot}
\end{figure}

Double-SOI structure succeeded to compensate radiation damages~\cite{DSOI}.
Method of recovery from radiation damage in this study that removing trapped charges from the BOX via tunneling is quite new.
Basically damaged detector will be replaced at the time of its lifetime in the experiments written above.
Double-SOI structure is also new approach for compensation of radiation damage.
But their approach is different in that it tries to cancel positive potential by applying negative voltage to middle SOI layer.
Fowler-Nordheim tunneling utilized in this study is one of the tunneling which occur by applying a high electric filed around 5~MV/cm.
The electrons tunnel through triangular potential of $\mathrm{Si-SiO_2}$ junction as shown in Figure~\ref{fig:fn}.
Current of FN tunneling express by
\begin{equation}
  J=A\sEox\exp\left(-\frac{B}{\Eox}\right)
  \label{eq:fncurrent}
\end{equation}
here $A$ and $B$ are as written below.
\begin{eqnarray}
  A & = & \frac{e^3}{8\pi h\phi_{\mathrm B}(m^{*}_{\mathrm{ox}}/m_0)}\\
  B & = & \frac{8\pi\sqrt{2m^*_{\mathrm{ox}}}\phi_B^{3/2}}{3he}
  \label{eq:AB}
\end{eqnarray}
FN current is proportional to the square of the electric field across the oxide layer~\Eox.
Therefore FN current is very sensitive to supplied voltage to oxide layer compared with other tunneling.
If we transform Equation~\ref{eq:fncurrent} and take the log of both side, we obtain this equation
\begin{equation}
  \ln\left(\frac{J}{\sEox}\right)=\ln(A)-\frac{B}{\Eox}.
  \label{eq:fnplot}
\end{equation}
The equation means the plot of \fnc\ as a function of \invE\ shows linear relationship as shown in Figure~\ref{fig:fnplot}~\cite{fnplot}.
High voltage that shows linearity in the FN plot will be able to remove trapped charges from the BOX.

\section{Experimental Procedure}

\begin{figure}[htbp]
  \centering
  \includegraphics[width=0.8\textwidth]{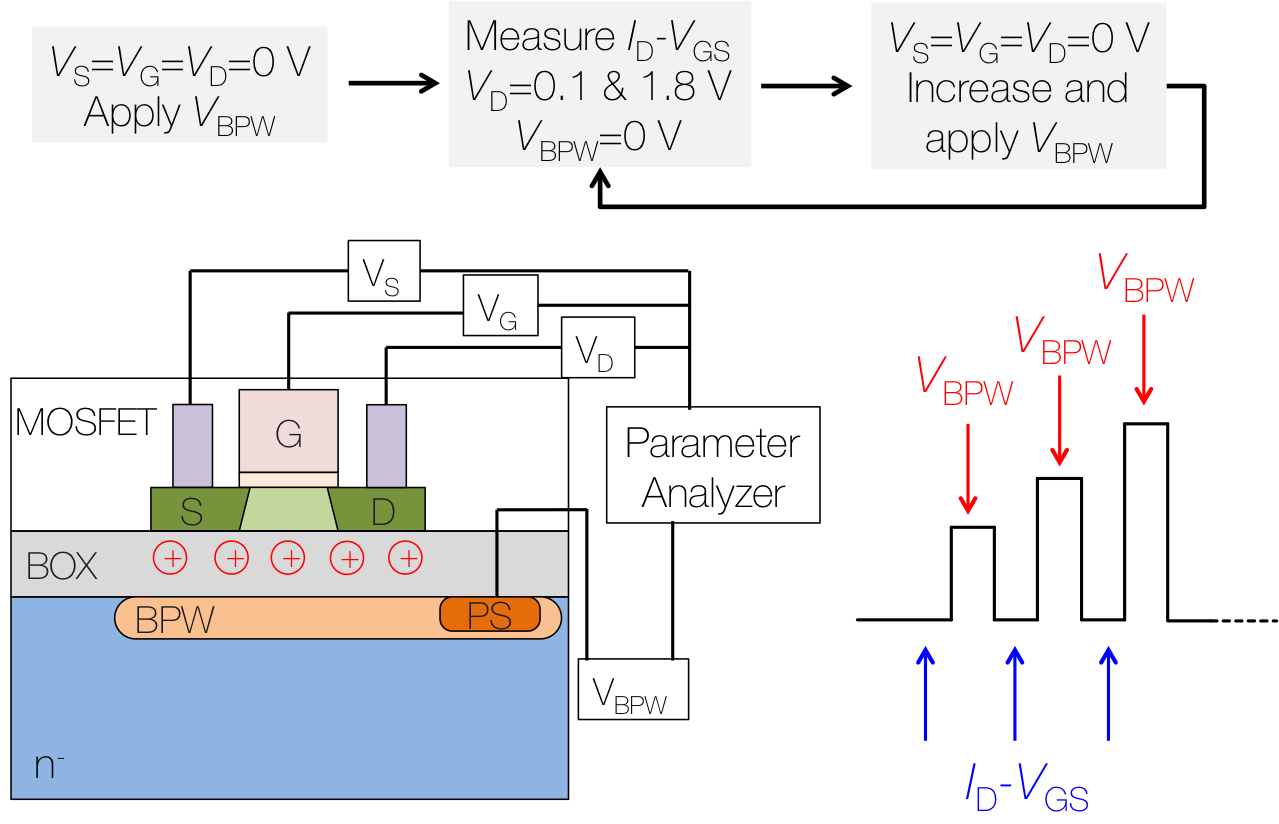} 
  \caption{Experimental setup and procedure for measurement of \idvg\ and applying \vbpw.}
  \label{fig:stress}
\end{figure}

A test element group of MOSFETs called TrTEG which has different gate width and length used for evaluation of characteristics of MOSFETs before and after irradiation.
TrTEG was fabricated using a 0.2 ~\um\ FD-SOI CMOS process technology for a quantum imaging application prepared by LAPIS Semiconductor Co., Ltd~\cite{lapis}.
The thickness of the BOX is $\Tbox=200$~nm.
Characteristics of MOSFETs, \idvg\ and \ileak\ measured with source meter units (SMU) manufactured by Keithley Instruments, Inc.. 
\xray\ irradiated to the TrTEG up to 250~Gy and 1~kGy at KEK.
The \xray\ source is the \xray\ tube of copper K$\alpha$ line of 8~keV.
The dose rate of the \xray\ was around 130~mGy/s.
All pads for MOSFETs were connected to the ground during the irradiation.
For evaluation of characteristics of MOSFETs, \idvg\ with $\vd=1.8$~V measured before and after irradiation.
Threshold voltage \vth\ and leak current \ileak\ derived from \idvg\ and used them to evaluate performances of MOSFETs.
Definition of \vth\ and \ileak\ are
\begin{eqnarray}
  \vth &=&\vgs\ \mathrm{at}\ \id=0.1\times W/L\ (\mu \mathrm{A})\\
  \ileak &=& \id\ \mathrm{at}\ \vgs=0~\mathrm{V}.
  \label{eq:vth}
\end{eqnarray}
To remove trapped charges from the BOX via FN tunneling, high voltage applied between the MOSFET and the BPW as shown in Figure~\ref{fig:stress}.
Source, Gate and Drain are fixed to 0~V and the high voltage up to 150~V is supplied by the BPW (\vbpw) for three seconds this time.
After applying \vbpw, \idvg\ with $\vd=1.8$~V measured and then derived \vth\ and \ileak. 
FN plot can obtain from applied \vbpw\ assuming $\Eox=\vbpw/\Tbox$ and measured \ibpw\ which is a current through BPW during \vbpw.

\section{Results and Discussion}
\subsection{Recovery of \vth}
\label{sec:vth}
\begin{figure}[htbp]
  \centering
  \includegraphics[width=0.8\textwidth]{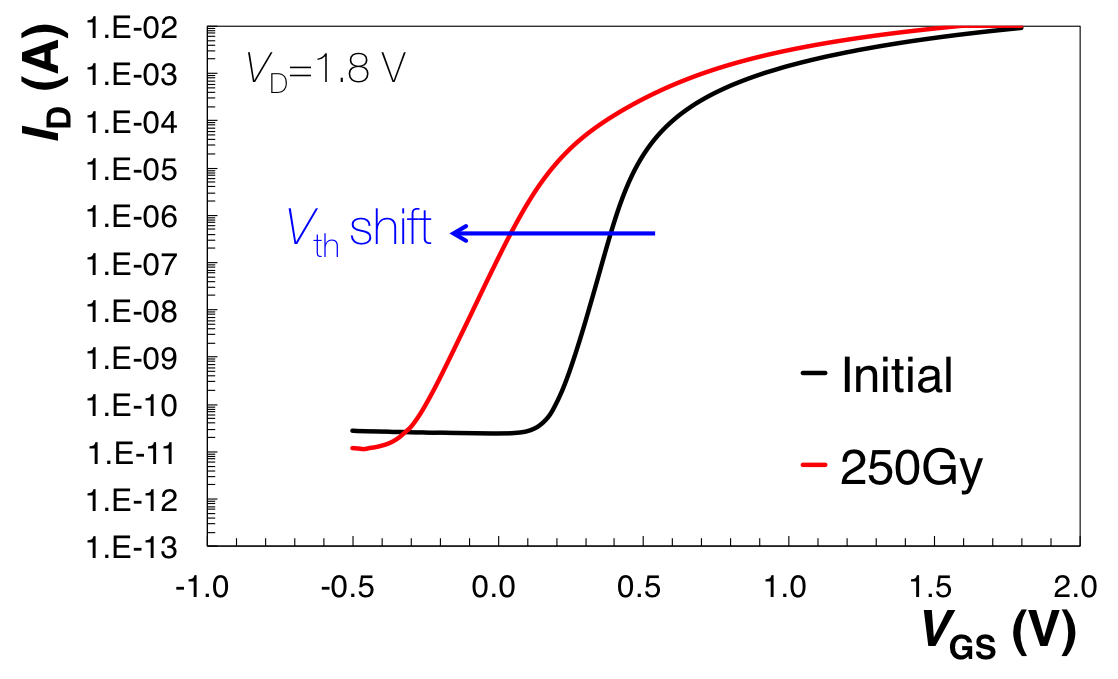}
  \caption{\idvg\ with $\vd=1.8$~V before and after irradiation of \xray\ of 250~Gy.}
  \label{fig:idvg250}
\end{figure}

\begin{figure}[htbp]
  \centering
  \subfigure[\ibpw]{
  \includegraphics[width=0.45\textwidth]{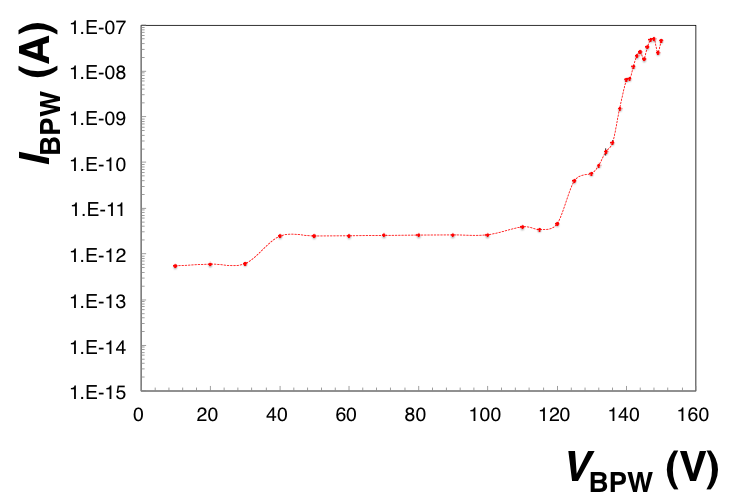}
  \label{fig:ibpw250}}
  \subfigure[FN plot]{
  \includegraphics[width=0.45\textwidth]{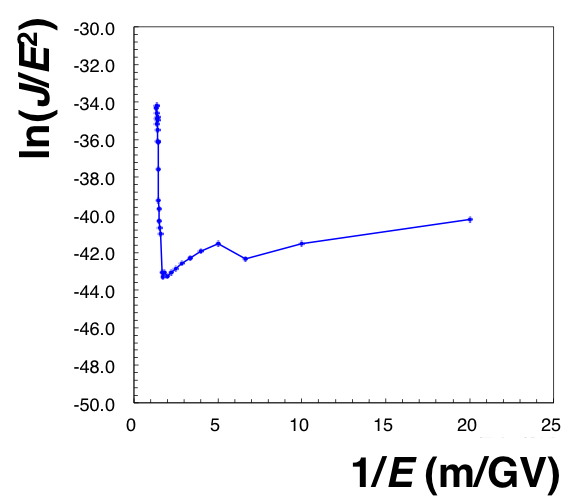}
  \label{fig:fnplot250}}
  \caption{(a) Current at the BPW \ibpw\ during applying \vbpw. \ibpw\ increases exponentially around $\vbpw=120$~V ($\Eox=6$~MV/cm). (b) FN plot during applying \vbpw\ up to 150~V. Linear relationship between \fnc\ and \invE\ can been seen $\vbpw>120$~V ($\invE<1.67$).}
  \label{fig:bpw250}
\end{figure}

\begin{figure}[htbp]
  \centering
  \includegraphics[width=0.8\textwidth]{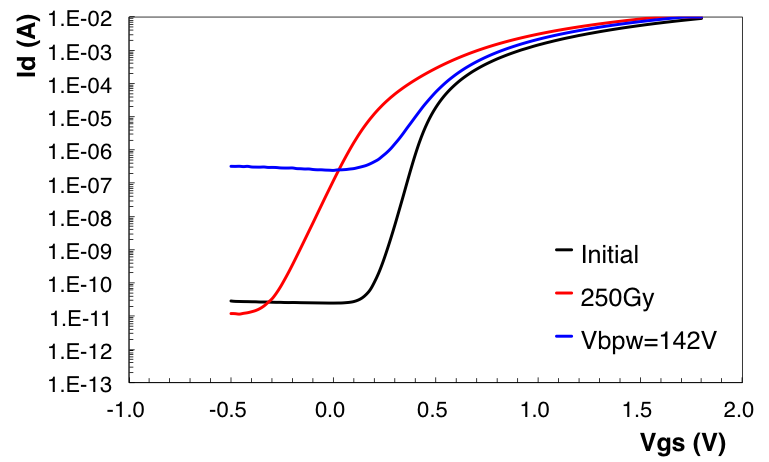}
  \caption{\idvg\ after applying $\vbpw=142$~V (blue line). Negative shift of \vth\ recovered close to pre-irradiation level but \ileak\ increased.}
  \label{fig:idvg142v}
\end{figure}

\begin{figure}[htbp]
  \centering
  \subfigure[Summary of \vth]{
  \includegraphics[width=0.45\textwidth]{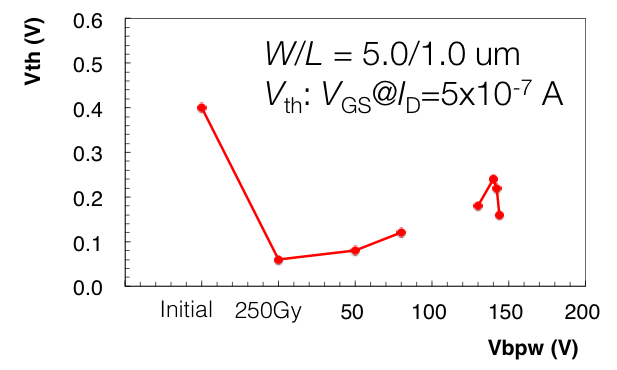}
  \label{fig:vth142v}}
  \subfigure[Summary of \ileak]{
  \includegraphics[width=0.45\textwidth]{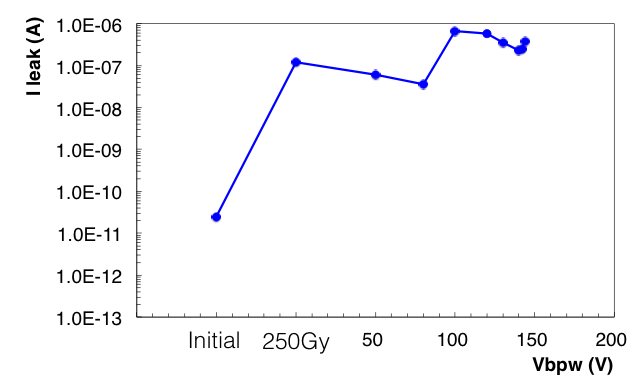}
  \label{fig:ileak142v}}
  \caption{Summary of \vth\ and \ileak\ before and after irradiation of \xray\ of 250~Gy, and after applying \vbpw\ up to 142~V.
\vth\ at region of $100\leq\vth\leq 120$ can not determine due to high leak current as shown in Figure~\ref{fig:ileak142v}.}
  \label{fig:summary142v}
\end{figure}

\idvg\ showed negative shift of \vth\ after the \xray\ irradiation of 250~Gy as shown in Figure~\ref{fig:idvg250}.
To recover this negative shift to the pre-irradiation level, \vbpw\ applied up to 150~V ($\Eox=7.5$~MV/cm).
Measured \ibpw\ in Figure~\ref{fig:ibpw250} increases exponentially around $\vbpw=120$~V.
In the FN Plot as shown in Figure~\ref{fig:fnplot250}, linearity can be seen where \vbpw\ is more than 120~V (6~MV/cm).
It means we observed FN tunneling from the BPW to the BOX at the electric filed is more than 6~MV/cm.
After applying \vbpw\ up to 142~V, \idvg\ shifted to the pre-irradiation level as shown in Figure~\ref{fig:idvg142v}.
Summary of \vth\ is shown in Figure~\ref{fig:vth142v}.
\vth\ at $\vbpw=100\sim120$~V can not determine due to high leak current.
\vbpw\ greater than 120~V where we can see the linearity in the FN plot recovered \vth.
It indicates that trapped charges decreased by injecting the electrons into the BOX via FN tunneling.
However \ileak\ increased after applying \vbpw\ as you can see \idvg\ and summary of \ileak\ shown in Figure~\ref{fig:ileak142v}.
\subsection{Recovery of \vth\ without increase of \ileak}
\begin{figure}[htbp]
  \centering
  \includegraphics[width=0.45\textwidth]{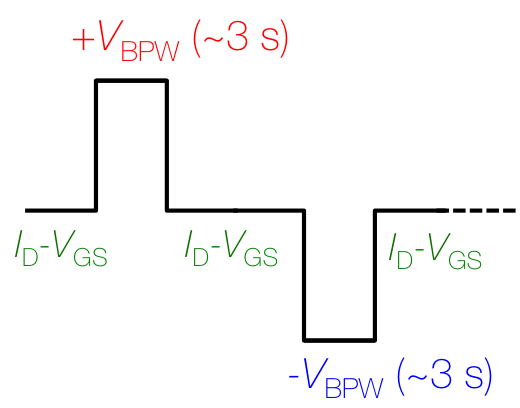}
  \caption{Experimental procedure for the recovery of \vth\ without increase of \ileak. +\vbpw\ is for recovery of \vth\ and -\vbpw\ is for suppression of \ileak\ due to +\vbpw.}
  \label{fig:pm140v}
\end{figure}

\begin{figure}[htbp]
  \centering
  \includegraphics[width=0.8\textwidth]{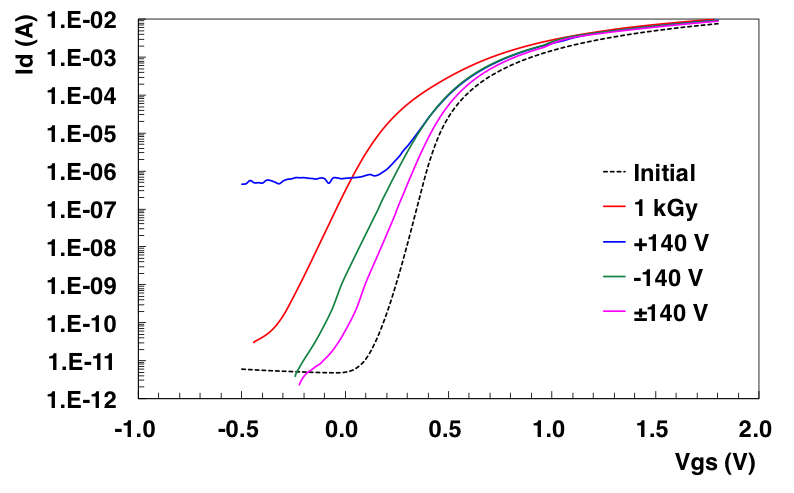}
  \caption{\idvg\ with $\vd=1.8$~V before and after irradiation of \xray\ of 1~kGy, and after repeating of $\pm\vbpw=140$~V.}
  \label{fig:idvg1k}
\end{figure}

\begin{figure}[htbp]
  \centering
  \subfigure[Summary of \vth]{
  \includegraphics[width=0.45\textwidth]{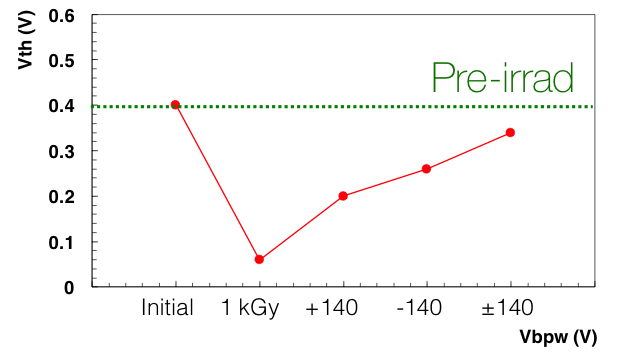}
  \label{fig:vth1k}}
  \subfigure[Summary of \ileak]{
  \includegraphics[width=0.45\textwidth]{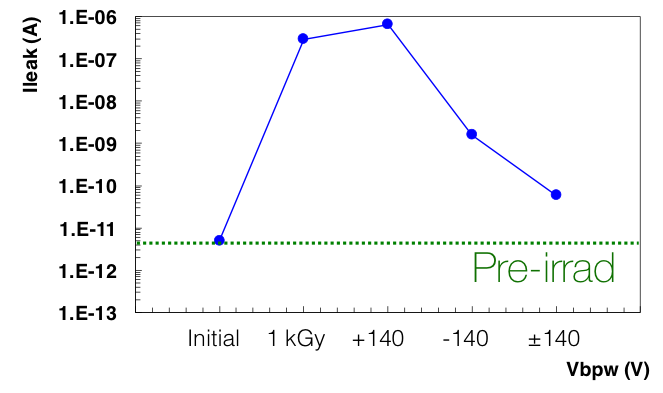}
  \label{fig:ileak1k}}
  \caption{Summary of \vth\ and \ileak before and after irradiation of \xray\ of 1~kGy, and after applying $\pm\vbpw=140$~V. \vth\ recovered close to pre-irradiation level without increase of \ileak.}
  \label{fig:summary1k}
\end{figure}

To suppress increased \ileak\ after applying \vbpw, negative \vbpw\ fixed to -140~V supplied to the BPW.
The sample irradiated 1~kGy used to clarify the effect of the irradiation and the \vbpw\ compared to the sample irradiated 250~Gy.
Applying $\vbpw=+140$~V and then $\vbpw=-140$~V repeated about twenty times to recover \vth\ and to suppress increased \ileak\ simultaneously.
\idvg\ measured after each $\pm\vbpw$.
Procedure after irradiation shows Figure~\ref{fig:pm140v}.
\idvg\ after repeating $\pm\vbpw=140$~V is shown in Figure~\ref{fig:idvg1k}, and summary of \vth\ and \ileak\ are shown in Figure~\ref{fig:vth1k} and~\ref{fig:ileak1k}.
\idvg\ after $\vbpw=+140$~V is blue line in the Figure~\ref{fig:idvg1k}.
Behavior of \idvg\ is the same as the Figure~\ref{fig:idvg250} and \ileak\ increased due to +\vbpw.
After $\vbpw=-140$~V shows green line and increased \ileak\ is suppressed to the pre-irradiation level.
Repeating twenty times of $\pm\vbpw=140$~V shows magenta line and it recovered close to the pre-irradiation level.
We succeeded to recover \vth\ close to the pre-irradiation level without increase of \ileak\ as shown in Figure~\ref{fig:vth1k} and~\ref{fig:ileak1k}. 
It means that +\vbpw\ recovered \vth\ via FN tunneling and $-\vbpw$\ suppressed increase of \ileak\ due to +\vbpw.

\section{Summary}
We succeeded to compensate radiation damage of negative shift of \vth\ via FN tunneling by applying high voltage between MOSFET and the BPW.
It suggest to be able to compensate the TID effects by simply applying high voltage to the BPW for a few seconds.
We will check other parameters of MOSFET such as \gm, {\it S} and so on related to performance of analog circuit (e.g. the small-signal gain).
The principle of increase of \ileak\ by +\vbpw\ will be investigated by using TCAD device simulation (HyENEXSS~\cite{hyenexss}).
The irradiation up to a few hundred kGy or MGy which are level of the inner most layer pixel detector of the LHC or the HL-LHC experiments is planning. 

The double-SOI, it have to supply negative voltage to middle SOI layer (VSOI2) continuously after irradiation and appropriate VSOI2 will change according to the dose.
It indicates they have to develop additional circuit to control VSOI2 automatically.
But our method using FN tunneling can recover performance of MOSFET by applying \vbpw\ a few seconds.
\vbpw\ is determined by the electric filed (around a few MV/cm) for FN tunneling and the thickness of oxide layer.
It means it does not depend on dose level.
However it is much better to have several way to compensate TID effects for future high energy physics experiments at extremely high radiation level.
Double-SOI and our method are promising.

\end{document}